# Summary and Conclusions of the 'JRA Beam Telescope 2025'-Forum at the 6th Beam Telescopes and Test Beams Workshop

*J. Dreyling-Eschweiler[1], H. Jansen[1] (eds.), M. S. Amjad[2], J.-H. Arling[1], T. Coates[3], A. Dätwyler[4], D. Dannheim[5], M. W. U. van Dijk[5], T. Eichhorn[1], A. Gerbershagen[5], O. Girard[6], B. Gkotse[5], F. J. Iguaz[7], J. Kroll[8], C. Nellist[9], F. Ravotti[5], E. Rossi[1], A. Rummler[5], F. Salvatore[3], S. Spannagel[5], M. Weers[10], J. Weingarten[10]*

## 1. Motivation

The "Beam Telescopes and Test Beams" (BTTB) workshop aims at bringing together the community involved in beam tests. It therefore offers a suitable platform to induce community-wide discussions on possible future common beam telescopes organised by means of a joint research activity (JRA). The input of the community to such a discussion is essential for determining the needs and requirements of a future common effort.

A previous workshop, the "Future Opportunities for Test Beams at DESY" workshop in October 2017 [1], evaluated the EUDET-type beam telescopes and started a discussion on future improvements and possible continuations. The outcomes were:

- The EUDET-type beam telescopes were recognized as "one of very few examples of beam telescopes that are used by many different users from various experiments".
- The community values the 'whole package' consisting of telescope hardware and telescope DAQ, DAQ software and reconstruction framework.
- A dedicated user support and continuous maintenance for each of the components are crucial.
- A common and standardized timing layer was suggested for a high rate capability and a precise hit timing.
- A common cold box was suggested.

The 'JRA Beam Telescope 2025' forum held at the 6th edition of the BTTB workshop series in January 2018 [2] picked up and continued this discussion focusing on beam telescope related user demands.

[1] Deutsches Elektronen-Synchrotron (DESY), Notkestr. 85, 22607 Hamburg, Germany
[2] Department of Physics and Astronomy, University College London (UCL), Gower Street, London, WC1E 6BT, United Kingdom
[3] University of Sussex, Sussex House, Falmer, Brighton, BN1 9RH United Kingdom
[4] Physik-Institut, University of Zurich, Winterthurerstrasse 190, 8057 Zurich, Switzerland
[5] European Organization for Nuclear Research (CERN), Route de Meyrin, 1217 Meyrin, Switzerland
[6] Laboratory for High Energy Physics (LPHE), Ecole polytechnique fédérale de Lausanne (EPFL), BSP - Cubotron, 1015 Lausanne, Switzerland
[7] Institut de Recherche sur les lois Fondamentales de l'Univers (IRFU), Université Paris-Saclay, 91191 Gif-sur-Yvette, France
[8] Institute of Physics of the Czech Academy of Sciences Na Slovance 1999/2, 18221 Prague 8, Czech Republic
[9] II. Physikalisches Institut - Kern- und Teilchenphysik, Georg-August-Universität Göttingen, Wilhelmsplatz 1, 37073 Göttingen, Germany
[10] Lehrstuhl für Experimentelle Physik IV, Technische Universität Dortmund, Otto-Hahn-Str. 4a, 44227 Dortmund, Germany

## 2. Status of EUDET-type beam telescope package

EUDET-type beam telescopes have seen continuous development over the last decade and have become a commonly used infrastructure. They enable a wide detector R&D program and provide hardware (sensors, mechanics, TLU), DAQ software (EUDAQ) and a reconstruction software (EUTelescope). Today, there are seven copies at five different beam lines offering the same measurement setup at different beams [3]. The evolution of this package and the performance was reviewed in a BTTB contribution [4] and again summarized for the forum [5]:

- **Telescope hardware**: Mimosa26 sensors provide an excellent spatial resolution and a low material budget, allowing their usage at beam energies available at DESY, ELSA or CERN PS. Their integration time amounts to 115.2 µs. The EUDET TLU provides a trigger signal and a trigger-busy-handshake communication with the DUT DAQs. This enables a synchronisation via trigger IDs. Movable stages provide a standardized mechanical interface for the integration of devices under test (DUT).

- **EUDAQ**: EUDAQ is a modular, lightweight, distributed, generic top-level DAQ software written in C++. Standardized user interfaces are in place for communication with the DUT DAQs: sending raw data, logging, and converting raw data for further analysis. EUDAQ provides a Finite State Machine (Non-initialised, Initialised, Configured, Run Started and Run Stopped, Error) with appropriate transitions (OnInitialise, OnConfigure, OnStartRun, OnEndRun) and an Online Monitor. EUDAQ Version 2 enables data taking in multiple data streams.

- **EUTelescope:** EUTelescope is a generic track reconstruction framework written in C++. It is based on the ILCsoft framework using Marlin processors and LCIO file format. Available processors are, among others: data conversion, hot channel masking, clustering, hit maker, alignment, and track reconstruction by different fit methods.

- **Organisation & User interaction:** The development was funded and is supported by European frameworks: EUDET, AIDA and nowadays AIDA2020 (WP15.2 and WP5 (EUDAQ)). DESY is playing a central role in the coordination of the user support and the user-driven developments. Many results mainly in detector R&D in HEP were achieved by using EUDET-type telescopes.

## 3. Forum Discussions and Conclusions

The discussion was well received and many different aspects even beyond the scope were brought up and reviewed by the participants. The status of the reconstruction framework 'EUTelescope' and the general requirements of a reconstruction framework were extensively discussed. A couple of near-future collaborative upgrade projects of the existing hardware have been proposed and a mandate was given out to substantiate the efforts on a new common infrastructure. The participants agreed that a final road-map should include a common funding and/or an umbrella organisation.

## 3.1 Hardware upgrades

The participants agreed that an initiative for a common effort should be started promptly, due to the time scale of a such an effort and the aging of the existing EUDET-type beam telescopes. Two different levels of actions were supported: short-term upgrades continuing the usage of Mimosa26 sensors and long-term developments based on a new sensor choice. It was concluded that a common beam telescope equipped with recent sensor technology needs substantial development time of the order of at least three years. Short-term upgrades for the existing Mimosa26-based telescopes are:

- Using EUDAQ2 data taking with the AIDA TLU (see 3.2.).
- Working on permanent installations of a fast (i.e. faster than Mimosa26) timing reference plane for DESY and CERN EUDET-type telescopes. Efforts for the existing integrations of the FEI4 (25 ns) and the Timepix3 (1.56 ns) were mentioned.
- Replacing the current NI-based MimosaDAQ by the MMC3 board [6].

For a new common beam telescope the development of the entire package depends on the decision of the sensor. The discussion covered different considerations:

- The sensor choice depends on beam conditions and the user needs. One difficulty for a common effort is to cover as many user needs as possible, likely requiring a compromise between them.
- The sensor should come with superior characteristics in comparison to the Mimosa26: thin (50 μm) in order to be usable at GeV beamlines, an excellent intrinsic resolution (< 3.5 μm), and considerably faster than 115 μs.
- 10 to 15 participants prefer a sensor with improved intrinsic resolution compared to the Mimosa26.
- Sensors with a larger area do not seem to be essential for R&D in tracking detectors. The demand for a Single Arm Large Area Telescope (SALAT), developed within AIDA, has been small to non-existent. This is different in test beams for calorimeters.
- The read-out mode (data-driven, frame based or rolling shutter) is important for the whole package design.
- A survey of available sensors and read-out systems should be carried out in order to find the then best sensor for a future common beam telescope. However, given that sufficient resources become available, a dedicated sensor development for a future common beam telescope could be carried out.

Apart from the sensor choice, important items include:

- telescope monitoring (including self-test)
- debug capabilities (JTAG, sensor currents)
- improved alignment along beam direction
- reliable automation for automated data taking.

## 3.2 EUDAQ status and development

The participants agreed that no crucial upgrade of the current EUDAQ framework is needed, since the scope and the upcoming development of EUDAQ cover (mainly) all needs. EUDAQ development should always comply with modularity, universality and extensibility.

The EUDAQ framework has a modular and top-level structure and is – due to continuous maintenance and by design – ready for any hardware upgrade. Version 2 together with the AIDA TLU provide backward compatibility to existing user integrations (one data collector or "EUDET mode") as well as added flexibility owing to the possibility of data taking in multiple data streams including a synchronisation by Trigger ID ("mixed mode"). Additionally, a new synchronisation concept by providing a common clock to the DUT DAQs ("AIDA mode") is available. Final tests, optimization and documentation are on the way, a release for the user community is foreseen for mid 2018.

## 3.3 EUTelescope status

The participants discussed extensively advantages and drawbacks of EUTelescope. Main outcomes of the participant's assessment and experience of EUTelescope were:

- EUTelescope has achieved a level of flexibility and various scopes which can be complicated, but many results were achieved by different groups in the past.
- The installation is difficult due to many dependencies.
- Documentation (incl. web pages) and support can or should be updated and improved.
- The LCIO-based data format and the MARLIN processor framework brings along an overhead and is maybe not the proper data format.
- Refurbishing the framework could be a solution, however this requires a substantial effort of experienced developers.

## 3.4 Requirements for a new common reconstruction framework

Requirements and scope for a common reconstruction framework targeting the analysis of beam telescope data were discussed. These include:

- Simple to start with/to get first results
- Modularity
- Documentation for more complicated analyses
- Covering many use cases
- Online DQM, maybe including some build-in quick reconstruction.

A new framework could be considerably more light-weight, documentation and user support could be well-organised from the start. However, its development requires substantial input from an experienced developer with an enduring commitment. It must then be supported by at least a critical fraction of the community. Alternative reconstruction frameworks have been developed by different groups. A possible extension of one of these to a commonly used framework would need to be reviewed, if desired.

## 3.5 Organisation

For a future collaborative effort the participants point out various structural possibilities. It was generally concluded that the summary of this forum should serve as a starting point for further implementation of an umbrella organisation covering coordinative tasks. Ideas and considerations were:

- Implementation of mailing list and regular meetings.
- Questionnaire to collect individual user wishes.
- The foundation of a collaboration similar to RD50 was mentioned, which would imply the definition of a strategy including commitments by the participating institutions.
- A new EU funding framework is not decided yet, the earliest start would be in 2021. Hence, a possible future umbrella organisation should be founded outside an EU framework, but with the possibility to acquire EU fundings when applicable.
- Users of a common infrastructure have to be convinced that an organisation and efforts by the users are critical for its success.

## 3.6 Miscellaneous

During the forum an additional aspect was shortly discussed: An overview of available beamlines, their infrastructure and contacts is needed. It was mentioned that Christoph Rembser has compiled such an overview, however the idea came up to implement an online database, similar to the one set up for the irradiation facilities within the AIDA2020 framework [7, 8]. This is possible, but it has to be organised that the facility managers are responsible for their entry.

# 4. Summary

The forum and its discussions were well received and the participants concluded that appropriate actions should be undertaken promptly. Specific hardware and software proposals were discussed, with an emphasis on improving current common EUDET-type telescopes based on Mimosa26 sensors towards higher trigger rate capabilities in convolution with considerably improved time resolution. EUDAQ as a common top level DAQ and its modular structure is ready for future hardware. EUTelescope fulfils many requirements of a common reconstruction framework, but has also various drawbacks. Thus, requirements for a new common reconstruction framework were collected. A new common beam telescope evolves with the sensor decision and the whole package including a reconstruction framework depends on that decision.

# References

[1] Indico Pages of FOTB at DESY: https://indico.desy.de/indico/event/17998/
[2] Indico Pages of BTBB 6 in Zurich: https://indico.desy.de/indico/event/18050/
[3] Web portal, Hardware description and user manual of EUDET-type telescopes: http://telescopes.desy.de
[4] Talk by J. Dreyling-Eschweiler: https://indico.desy.de/indico/event/18050/session/9/contribution/11/material/slides/0.pdf
[5] Forum slides: https://indico.desy.de/indico/event/18050/session/11/material/0/0.pdf
[6] Talk by Y. Dieter: https://indico.desy.de/indico/event/18050/session/9/contribution/17/material/slides/0.pdf
[7] Talk by B. Gkotse: https://indico.desy.de/indico/event/18050/session/7/contribution/26/material/slides/0.pptx
[8] Link to the database of the irradiation facilities: http://irradiation-facilities.web.cern.ch/